\documentstyle[prd,aps,epsf]{revtex}
\begin{document}

\draft

\title{Classical Behaviour After a Phase Transition:\\
             I. Classical Order Parameters}

            \author{F. \ C.\ Lombardo$^{1,2}$ \thanks{f.lombardo@ic.ac.uk}, 
R.\ J.\ Rivers
            $^{1}$\thanks{r.rivers@ic.ac.uk}
            and F. \ D.\ Mazzitelli$^{2}$\thanks{fmazzi@df.uba.ar}}

\address{{\it
$^1$ Theoretical Physics Group; Blackett Laboratory, Imperial
College,
London SW7 2BZ\\
$^2$ Departamento de F\'\i sica, Facultad de Ciencias Exactas y Naturales\\
Universidad de Buenos Aires - Ciudad Universitaria,
Pabell\' on I\\
1428 Buenos Aires, Argentina}}

\maketitle

\begin{abstract}
We analyze the onset of classical field configurations after a
            phase
            transition. Firstly, we motivate the problem by means of
            a toy model in quantum mechanics. Subsequently, we consider a
            scalar field theory
            in which the system-field interacts with its environment,
            represented both
            by further scalar fields and by its own short-wavelength modes.
            We show that, for very rapid quenches, the order parameter can be
            treated classically
            by the time that it has achieved its ground state values
            (spinodal time).
\end{abstract}

%\vskip2pc \pacs{Pacs: 03.70.+k, 05.70.Fh, 03.65.Yz}
%\maketitle2
%\narrowtext
%\newpage

 \section{Introduction}
            The emergence of classical behaviour from a quantum system is a
            problem of interest in many branches of physics \cite{decogral},
            from the foundations of quantum mechanics, condensed matter and
            quantum optics, to quantum computing and quantum field theory. In
            the last few years, experimental evidence supports our theoretical
            understanding of the quantum to classical transition and also
            opens new avenues in the development of essential tools to
            understand the more hidden mysteries of the quantum world
            \cite{exp}.

            The quantum to classical transition, for a point particle, say,
            involves two different but
            very related conditions. The first one is that there should be
            {\it correlations}, i.e., the coordinates
            and momenta of a particle should be correlated in phase space,
            according to the classical equations of motion. For example, the
            Wigner function should have a peak at the classical trajectories.
            The second and equally important condition is the elimination of
            quantum interference between these classical trajectories, i.e.,
            {\it decoherence}. Once decoherence eliminates interference terms
            (absent in the `classical world'), the Wigner function becomes a
            good candidate for a classical probability
            distribution. The generalisation from particles to
            fields is straightforward, in principle, with the same
            attributes of correlations and decoherence.

            The onset of classical behaviour is a natural consequence of a
            quantum open system, triggered by the interaction between the
            system of interest and its environment. The coupling strength
            between system and bath sets the time-scale after which we can
            consider our system as classical, according to both of the
            conditions we mentioned before. This temporal scale is usually
            called the {\it decoherence time} $t_{\rm D}$. After this
            decoherence time we do not have macroscopic states in coherent
            superpositions anymore, and a probability distribution can be
            extracted that evolves by means of a generalised Fokker-Planck
            equation.

            Our concern in this paper will be the quantum to classical
            transition of a single scalar order-parameter field during
            continuous transitions,
            with the simplest double-well potential.  This is an idealisation
            of the
            phase transitions that are expected to occur at
            the GUT and EW scales in the
            standard big-bang cosmological model \cite{Kolb}.

            An analysis of phase transitions in quantum
            field theory that takes the non-equilibrium nature of the dynamics
            into account from first  principles  has
            only recently begun to be addressed. In particular, the naive
            picture of a
            classical order parameter (inflaton or Higgs) field $\phi$ rolling
            down an adiabatic effective potential, that was once a mainstay
            of cosmological field theory modelling, has been shown to be
            suspect \cite{cormier}. Alternatively, the suggestion by
            Kibble\cite{Kibble} that, while a non-adiabatic approach is
            essential, causality alone can set saturated bounds on time and
            distance scales during a transition, has been shown to be only
            partly true.
            The onset of classical behaviour is absolutely crucial in the
            work of
            Ref.\cite{Kibble} in that the experimental signal for the causal
            bound is the production of classical defects at a prescribed
            density. Topological defects are inevitable in most transitions,
            and they may have played a fundamental role in the formation of
            large scale structure (strings) \cite{zurek}. Moreover,
            superabundance of some topological defects may contradict the
            observational evidence (magnetic monopoles).

            In all the above-mentioned examples there is an order parameter
            which evolves from the false to the true vacuum of the theory: the
            Higgs fields in GUT and EW phase transitions, the inflaton
            field(s) in inflationary models, etc. Although these are quantum
            scalar fields with vanishing mean value (due to the symmetry of
            the initial quantum state), the order parameter is usually treated
            as a classical object. Our aim in this paper is to justify this
            assumption.

            Our approach follows the analysis started by two of us in
            Ref.\cite{lombmazz}, where we studied the emergence of classical
            inhomogeneities from quantum fluctuations for a self-interacting
            quantum scalar field. We have investigated there the decoherence
            induced on the long-wavelength field modes by coarse-graining the
            field modes with wavelength shorter than a critical value, in
            order to show how the system becomes classical due to the
            interaction with its environment (in that case composed of the
            short-wavelength field modes of the same field). For phase
            transitions the classicality
            of the order parameter can be analyzed along the same lines by
            extending the  model to accommodate spontaneous symmetry breaking.

            This is a difficult problem because, as has been
            pointed out in the literature, and as we will stress in what
            follows, non-perturbative and non-Gaussian effects are relevant in
            the analysis of phase transitions. As a
            trial run for this analysis we begin by presenting a toy model in
            which we will study the spread of a wave packet initially centered
            around the local maximum of a double well potential, paying
            particular attention to the influence of the environment on the
            Wigner function and on the reduced density matrix. Only later will
            we will extend our results to quantum field theory phase
            transitions.

            The paper is organized as follows. In the next section we study the
            evolution of a wave packet initially
            centered on the top of a double well potential.
            We describe the exact numerical evaluation of the evolution
            of the wave packet. We show that, as the coupling between the
            system and
            the environment decreases, the decoherence time increases. Due to
            the nonlinearities of the potential, when the coupling vanishes
            there is
            no classical limit, not even classical correlations. In section III
            we analyze the full problem in quantum field theory. Here, the
            situation
            is somewhat different. Although, again, the decoherence time
            increases as the
            coupling to the environment decreases, so does the spinodal time,
            the
            time for the order parameter to achieve its classical ground-state
            values. Thus, however weak the coupling, decoherence can still
            occur {\it before}
            the field has fallen to its classical level.

It is important to say that the issue of how the system evolves into the
classical theory has been addressed in Refs.\cite{salman,mottola}. For some
models, it has been shown that classicality emerges as a consequence of
profuse particle creation. A non-perturbative large occupation number of
long-wavelength modes produces, on average, a diagonal density matrix. This
dephasing effect occurs at late times. Here we will consider a different
model in which classicality is an early time event.

            Finally,
            in section IV, we present our final remarks and conclusions.

            \section{Toy model: double well potential and environment}

            The simplest field theory that permits a phase transition is that
            of a single real scalar field $\phi$, with action
            \begin{equation}
            S[\phi ] = \int d^4x\left\{ {1\over{2}}\partial_{\mu}
            \phi\partial^{\mu} \phi + {1\over{2}}\mu^2 \phi^2 -
            {\lambda\over{4!}}\phi^4\right\} \label{1}
            \end{equation}
            with ($\mu^2 >0$) ${\cal Z}_2$ symmetry breaking. At high
            temperature the
            symmetry is restored.
             As we shall see
            later, a phase transition induced by a sudden temperature quench
            can be described by an effective field theory in which there is a
            change of sign in the mass term of the scalar field
            \begin{equation}
            S^{\rm eff}[\phi] = \int d^4x \big[{1\over{2}}\partial_\mu\phi
            \partial^\mu \phi - {1\over{2}}m^2(t) \phi^2 - {1\over{4!}}\lambda
            \phi^4\big],
            \label{1eff}
            \end{equation}
            where  $m^2(t) = M^2 > 0$  for $t < 0$ and
            $m^2(t)= - \mu^2$  for $t$ sufficiently positive. This change of
            sign
            in $m^2(t)$ breaks the global ${\cal Z}_2$ symmetry for positive
            $t$.

            Understanding this transition, for even such a simple system, is
            difficult. As a preliminary exercise, we start by considering a
            toy model composed of a particle, a quantum anharmonic oscillator
            (the `system'), linearly coupled to an environment composed of an
            infinite set of harmonic oscillators. The total classical action
            is given by

            \begin{eqnarray}
            S[x,q_n] &=& S_{\rm syst}[x] + S_{\rm env}[q_n] + S_{\rm
            int}[x,q_n]
            \nonumber \\
            &=& \int_0^t ds \left[{1\over{2}} M ({\dot x}^2 - \Omega_0^2(t)
            x^2 -
            {\lambda\over{4}} x^4)\right.
            \nonumber \\
            &+&\left. \sum_n {1\over{2}} m_n ({\dot q}_n^2 - \omega_n^2
            q_n^2)\right] - \sum_n C_n x q_n,
            \label{action}
            \end{eqnarray}
            where $x$ and
            $q_n$ are the coordinates of the particle and the oscillators,
            respectively. The quantum anharmonic oscillator is coupled
            linearly to each oscillator in the bath with strength $C_n$. In
            analogy with $m^2(t)$ of (\ref{1}) we consider the simplest
            possible
            case of an {\it instantaneous} quench, in which $\Omega_0^2(t)=
            \Omega_0^2$ when
            $t<0$ and $\Omega_0^2(t)=-\Omega_0^2$ when $t>0$. The unstable
            particle, coordinate $x$, has an initial thermal distribution at
            temperature $T$, for which $\langle {\hat x}\rangle = 0$.  For
            $t>0$ it finds itself on the unstable hump of the potential  and
            falls towards its stable minima at $\langle {\hat x}^2\rangle =
            2\Omega_0^2/\lambda$.

            \subsection{The environment}

            In the absence of an environment, and considering $\lambda =0$,
            this situation has been discussed
            in detail by Guth and Pi\cite{guthpi,kim}. In the presence of an
            environment the relevant objects for analyzing the quantum to
            classical transition in this model are the reduced density matrix,
            and the associated Wigner function
            \begin{equation}
            \rho_{\rm r} (x,x',t)= \int dq_n\,\, \rho (x,q_n,x',q_n,t),
            \nonumber
            \end{equation}
            and
            \begin{equation}
            W_{\rm r} (x,p,t)= {1\over{2\pi}} \int_{-\infty}^{+\infty} dy~
            e^{ipy} ~ \rho_{\rm r}(x+{y\over{2}}, x-{y\over{2}},t).
            \end{equation}
            The reduced density matrix satisfies a closed master equation.
            This has been evaluated by Hu-Paz-Zhang \cite{hpz} for the
            quantum Brownian motion problem with $\Omega_0^2$ positive.
            Following the same procedure, we
            can write a corresponding master equation for the unstable particle
            \cite{pazzurek} by
            replacing $\Omega_0$ by $i \Omega_0$ in the Hu-Paz-Zhang
            result
            \begin{eqnarray}i\partial_t \rho_{\rm r}(x,x',t)
            &=&\langle x\vert[H,\rho_{\rm
            r}]\vert x' \rangle \nonumber \\
            &-& i\gamma(t)(x-x')(\partial_x - \partial_{x'})
            \rho_{\rm r}(x,x',t)
            \nonumber\\
            &+& f(t)(x-x')(\partial_x + \partial_{x'}) \rho_{\rm r}(x,x',t)
            \nonumber \\&-&
            iD(t)(x-x')^2\rho_{\rm r}(x,x',t). \label{meqbm}
            \end{eqnarray}
             In (\ref{meqbm}), $H={H}_{\rm syst}-{1\over{2}}M{\tilde
            \Omega}^2(t)$ where $\tilde\Omega^2(t)$
             renormalizes the natural frequency of the
            particle, $\gamma (t)$ is the dissipation coefficient, and $D(t)$
            and $f(t)$ are the diffusion coefficients, which produce the
            decoherence effects. They depend on the properties of the
            environment as
            \begin{eqnarray}
            \tilde\Omega^2(t) &=& -{2\over M}\int_0^t dt'\cosh(\Omega_0 t')
            \eta(t') \nonumber
            \\
             \gamma(t) &=& -{1\over 2M\Omega_0}\int_0^t
            dt'\sinh(\Omega_0 t') \eta(t')\nonumber
            \\
            D(t) &=& \int_0^t dt'\cosh(\Omega_0 t') \nu(t')\label{coef} \\
            f(t) &=& -{1\over M\Omega_0}\int_0^t dt'\sinh(\Omega_0 t')
            \eta(t'),\nonumber
            \end{eqnarray}
            where $\eta (t)$ and $\nu (t)$ are the dissipation and noise
            kernels, respectively,
            \begin{eqnarray}\eta (t)& =& \int_0^\infty d\omega I(\omega )
            \sin \omega t
            \nonumber \\
            \nu (t) &=& \int_0^\infty d\omega I(\omega ) \coth {\beta
            \omega\over{2}} \cos \omega t\nonumber, \nonumber\end{eqnarray}
            and $I(\omega )= {\cal O}(C_n^2)$ is the spectral density of the
            environment.

            The first term on the RHS of Eq.(\ref{meqbm}) gives the usual
            Liouville-like evolution; the term proportional to $\gamma$
            produces dissipation ($\gamma$ is the relaxation coefficient). The
            term proportional to the diffusion coefficient $D(t)$, which is
            proportional to $(x-x')^2$ and positive definite, gives the main
            contribution to the decoherence since it produces a
            diagonalization of the reduced density matrix. Let us write the
            reduced density matrix as
            \begin{equation}
            \rho_{\rm r}[x,x';t] = G[x,x',t] \exp{\Big[ - (x-x')^2 \int_0^t
            D(s)~ ds\Big]} .\label{qbmdecay0}\end{equation} Inserting this
            expression into the master equation it is easy to to see that the
            differential equation for $G[x,x',t]$ contains the usual
            Liouville-term plus additional contributions proportional to
            $D,\gamma$, and $f$. However, none of these additional terms is
            imaginary with the right sign for diffusion. An approximate
            solution of Eq.(\ref{meqbm}) is therefore \cite{zurekTA}
            \begin{equation}
            \rho_{\rm r}[x,x';t] \approx \rho^{\rm u}_{\rm r}[x,x',t]
            \exp{\Big[-(x-x')^2 \int_0^t D(s) ~ ds\Big]}
            ,\label{qbmdecay}\end{equation} where $\rho^{\rm u}_{\rm r}$ takes
            into account the unitary evolution.

            Alternatively, one can derive the following evolution equation for
            the reduced Wigner function of the system \cite{phz}:

            \begin{eqnarray}\dot{W_{\rm r}}(x,p,t)&=&\{H_{\rm syst},
            W_{\rm r}\}_{\rm PB}
            - {\lambda\over{4}} x
            \partial^3_{ppp}W_{\rm r}
            +2 \gamma (t) \partial_p(pW_{\rm r})\nonumber \\
            &+& D(t) \partial^2_{pp}W_{\rm r}
            - f(t) \partial^2_{px}W_{\rm r}.
            \label{fokker2}
            \end{eqnarray}
            Let us concentrate on the evolution equation (\ref{fokker2}). The
            first term on the right-hand side of Eq.(\ref{fokker2}) is the
            Poisson bracket, corresponding to the usual classical evolution.
            The second term includes the quantum correction (we have set
            $\hbar = 1$). The last three terms describe dissipation and
            diffusion effects due to coupling to the environment. In order to
            simplify the problem, we consider a high-temperature ohmic
            ($I(\omega )\sim \omega$) environment. In this approximation the
            coefficients in Eq.(\ref{fokker2}) become constants: $\gamma (t) =
            \gamma_0$, $f\sim 1/T$, and $D = 2 \gamma_0 k_{\rm B}T$. The
            normal diffusion coefficient $D$ is the term responsible for
            decoherence effects and at high temperatures is much larger than
            $\gamma_0$ and $f$. Therefore, in Eq.(\ref{fokker2}), we may
            neglect the dissipation and the anomalous diffusion terms in
            comparison to the normal diffusion. As we saw in (\ref{qbmdecay}),
            it is the diffusion exponential in
            this equation that enforces the approximate diagonalisation of
            ${\hat\rho}$ is this coordinate basis.

            It is important to note that this high-temperature approximation
            is well defined after a time-scale of the order of $1/(k_{\rm
            B}T)\sim\gamma_0/D$ (with $\hbar = 1$). The relevant regime of
            evolution for our systems takes place at times comfortably larger
            than this time-scale.

            \subsection{Numerical Analysis}

            We have solved  equation (\ref{fokker2}) numerically, in the
            high temperature limit, for different values of the diffusion
            coefficient $D$, in order to illustrate its relevance in the
            quantum to classical transition. Details are given
            elsewhere\cite{diana}.  We have chosen as initial condition a
            Gaussian state centered at $x_0=p_0=0$ with minimal uncertainty
            ($\sigma_x^2 = 0.5$ and $\sigma_p^2 = 0.5$). The Wigner function
            is initially positive definite, and  different from zero only near
            the top of the potential. We have set the coupling constant
            $\lambda = 0.01$, the renormalized frequencies ${\tilde \Omega} =
            \omega_n = 1$ (we are measuring time in units of ${\tilde
            \Omega}$) and the bare masses also equal to one.

            It is illustrative to examine first the exact result when the
            environment is absent (for this case we have solved numerically
            the Schr\"odinger equation). The initially Gaussian Wigner
            function begins to squeeze in the $x=p$ direction and, before the
            spinodal time ($t_{\rm sp}\sim 2.3$) it becomes a non-positive
            function (Fig. 1). Our definition of the spinodal time is that
            time at which $\langle {\hat x}^2\rangle_t = 2\Omega_0^2/\lambda$,
            its symmetry-breaking value. During the evolution, the Wigner
            function covers all the phase space (Fig. 2) and it is clear that
            it is not possible to consider it as a classical probability
            distribution. Although we started with a special initial state
            (Gaussian  with minimum uncertainty), the non-linearities of the
            potential make the Wigner function a non-positive distribution.

            Let us now consider a coupling with an environment such that the
            normal diffusion coefficient is $D = 0.01$. As expected, the
            evolution of the Wigner function is similar to the previous one at
            early times (Figs. 3 and 4). However, as can be seen from Figs. 5
            and 6, at long times it becomes positive definite and peaked
            around the classical phase space.

            The effect of the environment is more dramatic for larger values of
            the diffusion coefficient.  In our last example, $D=1$, the
            quantum to
            classical transition
            takes place almost instantaneously, even before
            the quantum particle pass through the spinodal point (Figs. 7 and
            8).

            It is interesting to note that, as the diffusion coefficient
            grows, the amplitude of the Wigner function decreases. This is
            due to the fact that the decoherence increases with $D$. The
            reduced density matrix diagonalizes. As a consequence, its
            `Fourier transform', the reduced Wigner function, spreads out.

            Our numerical results show explicitly
            that the existence of the environment is
            crucial in the quantum to classical transition. The decoherence
            time
            depends on the temperature and the coupling between system and
            environment
            through the diffusion coefficient $D$ \cite{zurek84}.

            \section{Phase transitions in field theory}

            We can now tackle the full quantum $\phi$-field theory of
            (\ref{1}). A simplified analysis of the model has been given 
            elsewhere\cite{lmr2}.
            As before, the onset of classical behaviour is due to
            the environment. For the infinite degree of freedom quantum field,
            the field ordering after the transition begins is due to the
            growth in amplitude of its unstable long-wavelength modes. For
            these modes the environment consists of the short-wavelength modes
            of the field, together with all the other fields $\chi_{\rm a}$
            with which the $\phi$ inescapably
            interacts\cite{lombmazz}.  The inclusion of explicit
            environment fields $\chi_{\rm a}$ is both a reflection of the fact
            that a scalar field in isolation is physically
            unrealistic, as well as providing us with a systematic
            approximation scheme.

            Although the system
            field can never avoid the decohering environment of
            its short-wavelength modes, to demonstrate the effect of an
            environment
             we first consider the case in which it is taken to be composed
            only of the fields $\chi_{\rm a}$.
            The short-wavelength modes of the $\phi$ field will be considered
            afterwards. Specifically, we take the simplest scalar classical
            action
            \begin{equation}
            S[\phi , \chi ] = S_{\rm syst}[\phi ] + S_{\rm env}[\chi ] +
            S_{\rm int}[\phi ,\chi ], \label{quaction}
            \end{equation}

            \begin{eqnarray}
            &&S_{\rm syst}[\phi ] = \int d^4x\left\{ {1\over{2}}\partial_{\mu}
            \phi\partial^{\mu} \phi + {1\over{2}}\mu^2 \phi^2 -
            {\lambda\over{4!}}\phi^4\right\}, \nonumber \\
            &&S_{\rm env}[\chi_{\rm a} ] = \sum_{\rm a=1}^N\int d^4x\left\{
            {1\over{2}}\partial_{\mu}\chi_{\rm a}
            \partial^{\mu}
            \chi_{\rm a} - {1\over{2}} m_{\rm a}^2 \chi^2_{\rm a}\right\}, 
            \nonumber
            \\
            &&S_{\rm int}[\phi ,\chi ] = - \sum_{\rm a=1}^N\frac{g_{\rm a}}{8}
            \int d^4x \phi^2 (x) \chi^2_{\rm a} (x),
            \label{Sint}
            \end{eqnarray}
            where $\mu^2, m^2 >0$.

            We see that the quantum mechanics action of (\ref{action}) is, in
            large part, a simplified version of (\ref{quaction}).  To make
            any progress it is
            important that the environment be as simple as possible. To that
            end we have included
            no explicit self-interaction of the $\chi_{\rm a}$ fields, which
            interact through the $\phi$
            field as intermediary. However,
            there is one significant difference in that the interactions of
            the $\phi$-field with
            the environment in (\ref{Sint}) are {\it quadratic}, and not
            linear as in
            (\ref{action}).  The more conventional choice of a linear coupling
            to the environment was made in the previous section (and
            \cite{diana}) to give a model directly comparable to similar
            particle models with no symmetry breaking,
            for which much work has been done. Although it has been adopted for
            quantum field theories\cite{kim} a linear-linear coupling is
            inappropriate and we have taken the simplest quadratic-quadratic
            couplings in $S_{\rm int}$ for convenience. The extension to Yukawa
            couplings is straightforward, and will be given elsewhere.
            We are reminded that, for non-linear couplings like
            $x^{\rm n}q_{\rm i}^{\rm m}$ in the quantum mechanical problem, one
            expects the master equation to contain terms of the form
            $iD^{({\rm n},{\rm m})}(t)(x^{\rm n}-x'^{\rm n})^2\rho_{\rm r}$.
            We shall find a similar effect here.
            The non-linear coupling to the environment is crucial to our
            conclusions.

            Further, in our present model, the environment has a strong
            impact upon the system-field, but not vice-versa, whenever
            possible.
            The simplest way to implement this is to take a large number 
            $N\gg 1$
            of $\chi_{\rm a}$
            fields with comparable masses $m_{\rm a}\simeq \mu$ weakly
            coupled to the $\phi$,
            with $\lambda$; $g_{\rm
            a}\ll 1$.  Thus, at any step, there are $N$ weakly
            coupled environmental fields influencing the system
            field, but only one system field to back-react upon
            the explicit environment.

            \subsection{Initial conditions}

            Before we examine the model
            in detail, there are some general observations to be made about
            initial conditions, and the way in which the transition is
            implemented. Like any simple scalar theory the model displays
             a continuous transition at a
            temperature $T_{\rm c}$,
            \[
            T_{\rm c}^2=\frac {\mu^2}{\lambda + \sum
            g_{\rm a}}\gg \mu^2
            \]

            The environmental fields $\chi_{\rm a}$ reduce
            $T_{\rm c}$ and, in order that $T_{\rm c}^2\gg \mu^2$, we must
            take
            \[
            \lambda + \sum g_{\rm a} \ll 1.
            \]
              For order of magnitude
            estimations it is sufficient to take identical $g_{\rm a} =\bar
            g/\sqrt{N}$, whereby
            \[
            1\gg 1/\sqrt N\gg\bar g\simeq \lambda.
            \]
            {\it i.e.} the $\chi_{\rm a}$ 'tadpole' diagrams completely
            overwhelm the $\phi$ self-interaction tadpole diagram in
            generating the $\phi$ thermal mass.

            With $\eta = \sqrt{6\mu^2/\lambda}$ determining the
            position of the minima of the potential and the final
            value of the order parameter, this choice of coupling
            and environments gives the hierarchy of scales
            \[
            \mu^2\ll T_{\rm c}^2 =
            O\bigg(\frac{\eta^2}{\sqrt{N}}\bigg)\ll\eta^2,
            \]
            important in establishing a reliable approximation
            scheme.
            Further, with this choice the dominant hard loop contribution of
            the $\phi$-field to the $\chi_{\rm a}$ thermal masses is
            \[
            \delta m^2_T = O (\bar{g} T^2_{\rm c}/\sqrt{ N}) = O(\mu^2/N)\ll
            \mu^2.
            \]
            Similarly, the two-loop (setting sun) diagram which is the first
            to contribute to the
            discontinuity of the $\chi$-field propagator is of magnitude
            \[
            \bar{g}^2 T_{\rm c}^2/N  = O(g\mu^2/N^{3/2})\ll\delta m^2_T,
            \]
            in turn. That is, the effect of the thermal bath on
            the propagation of the environmental $\chi$-fields is ignorable.

            This was our intention in model-making; to construct
            an environment that reacted on the system field, but
            was not reacted upon by it to any significant extent.
             We stress that this is not a
            Hartree or large-N approximation of the type that, to date, has
            been the standard way to proceed\cite{mottola,boya} for a
            {\it closed} system.

            We shall assume that the initial states of the system and
            environment are both thermal, at a temperature $T_{0}>T_{\rm c}$.
            We then imagine a change in the global environment (e.g. expansion
            in the early universe) that can be characterised by a change in
            temperature from $T_{0}$ to $T_{\rm f}<T_{\rm c}$. That is, we do
            not
            attribute the transition to the effects of the environment-fields.
            On
            incorporating the hard thermal loop tadpole diagrams of the $\chi$
            (and $\phi$) fields in the $\phi$ mass term leads to the effective
            action for $\phi$ quasiparticles,
            \[
            S^{\rm eff}_{\rm syst}[\phi ] = \int d^4x\left\{
            {1\over{2}}\partial_{\mu} \phi\partial^{\mu} \phi - {1\over{2}}
            m_{\phi}^2(T_0) \phi^2 - {\lambda\over{4!}}\phi^4\right\}
            \]
            where $m_{\phi}^2(T_0)=-\mu^2 (1-T_0^2/T_c^2)= M^2>0$. As a
            result, we
            can take an initial factorised density matrix at temperature $T_0$
            of the form ${\hat\rho}[T_0] = {\hat\rho}_{\phi}[T_0]
            {\hat\rho}_{\chi}[T_0]$, where ${\hat\rho}_{\phi}[T_0]$ is
            determined by the quadratic part of $S^{\rm eff}_{\rm syst}[\phi ]$
            and ${\hat\rho}_{\chi}[T_0]$ by $S_{\rm env}[\chi_{\rm a} ]$. Yet
            again, the many $\chi_{\rm a}$ fields have a large effect on
            $\phi$, but
            the $\phi$-field has negligible effect on the $\chi_{\rm a}$.
            Provided the change in temperature is not too slow the exponential
            instabilities of the $\phi$-field grow so fast that the field has
            populated the degenerate vacua well before the temperature has
            dropped to zero. Since the temperature $T_{\rm c}$ has no
            particular
            significance for the environment field, for these early times we
            can keep the temperature of the environment fixed at $T_{\chi} =
            T_{0}={\cal O}(T_{\rm c})$ (our calculations are only at the level
            of orders of magnitude).

            Since it is the system-field $\phi$ field whose behaviour changes
            dramatically on taking $T_{\phi}$ through $T_{\rm c}$, in this
            paper we adopt an {\it instantaneous} quench  for $T_{\phi}$ from
            $T_0$ to $T_{\rm f}=0$ at time $t=0$, in which $m^{2}_{\phi}(T)$
            changes sign and magnitude instantly, concluding with the value
            $m_{\phi}^2(t)=-\mu^2, t>0$ that we cited in (\ref{1}).  An {\it
            instantaneous} quench is sufficient to demonstrate the
            rapidity with which the environment forces the system
            field to become classical.  Meanwhile, for
            simplicity the
            $\chi_{\rm a}$ masses are fixed at the common value $ m\simeq\mu$.

            \subsection{Tracing out the $\chi$ fields}

            At time $t>0$ the reduced density matrix $\rho_{{\rm
            r}}[\phi^+,\phi^-,t]=\langle\phi^+\vert {\hat\rho}_r (t)\vert
            \phi^- \rangle$ is now
            \[
            \rho_{{\rm r}}[\phi^+,\phi^-,t] = \int {\cal D}\chi_{\rm a} ~
            \rho[\phi^+,\chi_{\rm a} ,\phi^-,\chi_{\rm a} ,t],
            \]
            where $\rho[\phi^+,\chi^+_{\rm a} ,\phi^-,\chi^-_{\rm a},t]=
            \langle\phi^+ \chi^+_{\rm a}\vert {\hat\rho}(t) \vert \phi^-
            \chi^-_{\rm a}\rangle$ is the full density matrix. Since we would
            like to be able to distinguish between different classical
            system-field configurations evolving after the transition, we will
            only interested in the field-configuration basis for this reduced
            density matrix (in analogy with the quantum Brownian motion model
            of the previous section). The environment will have had the effect
            of making the system essentially classical once $\rho_{\rm r}(t)$
            is, effectively, diagonal. [Our earlier comments on dephasing
            remain
            valid, and we stress again that our understanding of what
            constitutes
            classical behaviour  is essentially different from the
            dephasing effects found in Refs. \cite{salman,mottola}).] Quantum
            interference can then be ignored and
            the system is said to have decohered. At the same time,
            we obtain a classical probability distribution from the diagonal
            part of $\rho_{\rm r}(t)$, or equivalently, by means of the
            reduced Wigner functional. For weak coupling there will be no
            'recoherence' at later times in which the sense of classical
            probability will be lost.

            Its temporal evolution is given by

            \[\rho_{\rm r}[\phi_{{\rm
            f}}^+,\phi_{{\rm f}}^-,t]
            = \int d\phi_{{\rm i}}^+ \int d\phi_{{\rm
            i}}^- J_{\rm r}[\phi_{{\rm f}}^+,\phi_{{\rm f}}^-,t\vert
            \phi_{{\rm i}}^+,\phi_{{\rm i}}^-,t_0] ~~\rho_{\rm r}[\phi_{{\rm
            i}}^+ \phi_{{\rm i}}^-,t_0],
            \]
            where $J_{\rm r}$ is the reduced evolution operator

            \begin{equation}
            J_{\rm r}[\phi_{{\rm f}}^+,\phi_{{\rm f}}^-,t\vert \phi_{{\rm
            i}}^+,\phi_{{\rm i}}^-,t_0] = \int_{\phi_{{\rm i}}^+}^{\phi_{{\rm
            f}}^+} {\cal D}\phi^+ \int_{\phi_{{\rm i}}^-}^{\phi_{{\rm f}}^-}
            {\cal D}\phi^- ~ e^{i\{S[\phi^+] - S[\phi^-]\}} F[\phi^+,\phi^-].
            \label{evolred}
            \end{equation}
            The Feynman-Vernon \cite{feynver} influence functional $F[\phi^+,
            \phi^-]$ is defined as

            \begin{eqnarray}
            &&F[\phi^+,\phi^-]= \int d\chi^+_{{\rm a i}} \int d\chi^-_{{\rm a
            i}} ~ \rho_{\chi}[\chi_{{\rm a i}}^+,\chi_{{\rm a i}}^-,t_0] \int
            d\chi_{{\rm a f}} \nonumber \\
            &\times & \int_{\chi^+_{{\rm a i}}}^{\chi_{{\rm a
            f}}}{\cal D}\chi^+_{\rm a} \int_{\chi^-_{{\rm  a i}}}^{\chi_{{\rm
            a f}}}
            {\cal D}\chi^-_{\rm a}
            \exp{\left(i \{S_{\rm env}[\chi^+_{\rm a} ]+S_{{\rm int}}
            [\phi^+,\chi^+_{\rm a} ]\right)}\nonumber \\
            &\times & \exp{\left(-i\{S_{\rm env}[\chi^-_{\rm a}] + S_{{\rm
            int}}[\phi^-,\chi^-_{\rm a}]\} \right)}. \nonumber
            \end{eqnarray}

            Beginning from this initial distribution, peaked around $\phi =
            0$, we follow the evolution of the system under the influence of
            the environment fields, with Hamiltonian determined from
            (\ref{quaction}). From the influence functional we define the
            influence action $\delta A[\phi^+,\phi^-]$ as
            \begin{equation}
            F[\phi^+,\phi^-] =\exp {i \delta A[\phi^+,\phi^-]}.
            \end{equation}

            The total coarse-grained effective action is
            \[
            A[\phi^+,\phi^-]= S[\phi^+] - S[\phi^-] + \delta
            A[\phi^+,\phi^-].
            \]
            We will calculate the influence
            action to one loop (two vertices) for $N$ large using closed-time
            path correlators. It is the imaginary part which contains the
            information about the onset of
            classical behaviour. If $\Delta ={1\over{2}}(\phi^{+2} -
            \phi^{-2})$, we find
            \begin{equation}
            {\rm Im} \delta A = - \frac{\bar g^2}{16} \int d^4x\int d^4y ~
            \Delta (x) N_{\rm q} (x,y) \Delta (y).
            \label{imaginarypartIA}
            \end{equation}
            In (\ref{imaginarypartIA}) $ N_{\rm q} (x-y) = {\rm Re} G_{++}^2
            (x,y)$
            is the noise (diffusion) kernel, where $G_{++}$ is the relevant
            closed-timepath
            correlator of the $\chi$-field at temperature $T_0$. Non-leading
            one-loop terms are smaller by a factor ${\cal O}(N^{-1/2})$.

            The first step in the evaluation of the master equation is the
            calculation of the density matrix propagator $J_{\rm r}$ from Eq.
            (\ref{evolred}). In order to estimate the functional integration
            which defines the reduced propagator, we perform a saddle point
            approximation
            \[
            J_{\rm r}[\phi^+_{\rm f},\phi^-_{\rm f},t\vert\phi^+_{\rm
            i},\phi^-_{\rm i}, t_0] \approx \exp{ i A[\phi^+_{\rm
            cl},\phi^-_{\rm cl}]},
            \]
            where $\phi^\pm_{\rm cl}$ is the solution of the equation of
            motion ${\delta Re A\over\delta\phi^+}\vert_{\phi^+=\phi^-}=0$
            with boundary conditions $\phi^\pm_{\rm cl}(t_0)=\phi^\pm_{\rm i}$
            and $\phi^\pm_{\rm cl}(t)=\phi^\pm_{\rm f}$. It is very difficult
            to solve this equation analytically. For simplicity, we assume
            that the system-field contains only one Fourier mode with $\vec k
            = \vec k_0$. We are motivated in this by the observation that the
             long-wavelength modes start
            growing exponentially as soon as the quench is
            performed and rapidly dominate the fluctuation power
            spectrum\cite{boya}.
            Modes with $\vert k_0\vert^2
            > \mu^2$ will oscillate.

            Further, we are interested primarily in the
            order-parameter

            \[
            \phi (t)= \phi (k_0 = 0,t) =\lim_{V\rightarrow\infty}
\frac{1}{V}\int_{{\bf
            x}\in V} d^3x\,\phi({\bf x},t),
            \]
            and all our subsequent calculations will be for theis 
zero-frequency
            mode. We write the spatially-constant  classical solution as
            $\phi (s) =  f(s,t)$
            where $f(s,t)$ satisfies the boundary conditions $f(0,t)=
            \phi_{\rm i}$
            and $f(t,t) = \phi_{\rm f}$. Qualitatively, $f(s,t)$
            grows exponentially with $s$ for $t\leq t_{\rm sp}$, and oscillates
            for $ t_{\rm sp}<s<t$ when $t>t_{\rm sp}$.
            We shall therefore approximate its time dependence
            for $t\leq t_{\rm sp}$ as
            \begin{equation}
            f(s,t) = \phi_{\rm i} u_1(s,t) + \phi_{\rm f} u_2(s,t),
            \end{equation}
            where
            $u_1(0,t) = 1$, $u_1(t,t) = 0$ and
            $u_2(0,t) = 0$, $u_2(t,t) = 1$, with solution
            \[
            u_1(s,t) =  {\sinh[\mu (t - s)]
            \over{\sinh(\mu t)}},\,\,u_2(s,t)=  {\sinh(\mu s) \over
            {\sinh(\mu t)}}. \,\,\,
            \]
            We shall justify the use of the linear equation later.
            In fact, we shall see then that it is not an
            unreasonable approximation until almost $t_{\rm sp}$.

            In order to solve the master equation we
            must compute the final time derivative of the propagator $J_{\rm
            r}$, and after that eliminate the dependence on the initial field
            configurations $\phi^\pm_{\rm i}$ coming from the classical
            solutions $\phi^\pm_{\rm cl}$. This is the usual procedure, see
            Ref.\cite{lombmazz}.

            As we are solely interested in the onset of classical behaviour,
            it is sufficient to calculate the correction to the usual unitary
            evolution coming from the noise kernel. For clarity we drop the
            suffix ${\rm f}$ on the final state fields. If $\Delta =
            (\phi^{+2} - \phi^{-2})/2$ for the {\it final} field
            configurations, then the master equation for
            $\rho_r(\phi^+,\phi^-, t)$ is

            \begin{equation}
            i {\dot \rho}_{\rm r} = \langle \phi^+\vert [H,{\hat\rho}_{\rm r}]
            \vert \phi^-\rangle - i\frac{\bar g^2}{16} V \Delta^2 D( t)
            \rho_{\rm r}+ ... \label{master}
            \end{equation}
            The presence of $\Delta$ in (\ref{master}), rather than
$\phi_+ - \phi_-$, is a consequence of the quadratic coupling to the
            environment in $S_{\rm int}$.
            Since our main concern is with the diagonalisation of
            $\rho_{\rm r}$,
            it is not necessary to consider the evolution equation for the
            Wigner functional, as has been shown in the literature of quantum
            Brownian motion \cite{phz,diana}. The volume factor $V$ that
            appears in the master equation is due to the fact we are
            considering a density matrix which is a functional of two
            different field configurations, $\phi^\pm (\vec x )= \phi^\pm$,
            which are spread over all space. The time
            dependent diffusion coefficient $D_{\chi}(t)$ due to each of the
            many
            external environmental $\chi$ fields
            is given by

            \begin{equation}
            D_{\chi}( t)=
             3 \int_0^t ds ~ u(s,t) ~ {\rm Re}G_{++}^2(0; t-s),
            \label{diff}
            \end{equation}
            where

            \begin{equation}
            u(s,t)=\bigg[u_2(s,t) - \frac{{\dot u}_2(t,t)}{{\dot
            u}_1(t,t)}u_1(s,t)\bigg]^{2}.
            \label{u}
            \end{equation}
            For the case in hand of an instantaneous quench, $u(s,t) =
            \cosh^2\mu (t- s)$ when $t \leq t_{\rm
            sp}$, and is an oscillatory function of time when $t>t_{\rm sp}$.

            Although $G_{++}$ is oscillatory at all times,  the exponential
            growth of $u(t)$ enforces a similar behaviour on $ D_{\chi}(t)$.
            In the high temperature limit ($k_{\rm B}T \gg \mu$), the
            explicit expression
            for that contribution to the diffusion coefficient due to the
            $\chi_{\rm a}$ fields alone  is
            \begin{equation}
            D_{\chi}(t)={5(k_{\rm B}T)^2\over{64\pi^2}} \int_0^tds~ u(s,t)
            \int_0^\infty dp~{p^2\over{\omega^4}} \cos[2\omega s], \label{D}
            \end{equation}
            where $\omega^2 = p^2 + m^2$ and we have set $m^2 = \mu^2$.

            For times $\mu t \gg 1$, the integration in (\ref{D}) is
            dominated by the behaviour at $s=0$:
            \begin{equation}
            D_{\chi}(t)\sim (k_{\rm B} T_0/\mu)^2~ u(0,t)
            \sim (k_{\rm B} T_0/\mu)^2~ \exp [2\mu t].
            \label{Dt}
            \end{equation}

            The spinodal time $t_{\rm sp}$ is again defined as the time for
            which
            $\langle \phi^2\rangle_t \sim \eta^2= 6\mu^2/\lambda$. For
            $t>t_{\rm sp}$ the diffusion coefficient stops growing, and
            oscillates around $D(t=t_{\rm sp})$.

            \subsection{Short wavelength modes}

            In our present model the environment fields $\chi_{\rm a}$ are not
            the only decohering agents. The environment is also constituted by
            the short-wavelength modes of the self-interacting field $\phi$.
            Therefore, we split the field as $\phi = \phi_< + \phi_>$, where
            the system-field $\phi_<$ contains the modes with wavelengths
            longer than the critical value $\mu^{-1}$, while the bath or
            environment-field contains wavelengths shorter than $\mu^{-1}$.
            This gives an additional one-loop contribution $D_{\phi}(t)$ to
            the diffusion
            function with
            the same $u(s,t)$ but a $G_{++}$ constructed form the
            short-wavelength modes of the $\phi$-field as it evolves from the 
top of
            the potential hill. Without the additional powers of $N^{-1}$ to
            order contributions summation of loop diagrams is
            essential to get a reliable $G_{++}$. However, it is
            not necessary to calculate $D_{\phi}(t)$ in order to
            get a good estimate of $t_D$. Since the contribution
            of $D_{\phi}(t)$
            to the overall diffusion function is positive we can derive an
            {\it upper} bound on the decoherence time
            $t_D$ from the reliable diffusion functions $D_{\chi}$ alone.

            However, we would not expect the inclusion of the
            $\phi$-field to give a qualitative change.
            Specifically, we note that, if we take $\phi$-field propagators
            dressed by
            only the simplest tadpole diagrams, and ignore two-loop
            self-interaction diagrams then the
            diffusion correction due to the $\phi$ loop is now similar to that
            of the $\chi$ loops.  The effect is that the short-wavelength
            modes in the one-loop diagrams from which they are calculated have
            been kept at the initial temperature $T_0$, on the grounds that
            passing through the transition quickly has no effect on them.  The
            quench mimics a slower evolution of temperature in which only the
            long-wavelength modes show instabilities that the transition
            induces. That is, with ${\bar g}\simeq\lambda$ and no $1/N$ factor,
            the short-wavelength $\phi$
            modes would have the same effect on the dissipation,
            qualitatively, as {\it all} the explicit
            environmental fields put together. However, at an order of
            magnitude level there is no change, since the effect is to
            replace ${\bar g}^2$ by ${\bar g}^2 + O(\lambda^2) = O({\bar
            g}^2)$.

            \subsection{The decoherence time}

            Using the positivity of $D_{\phi}$ we estimate the
            decoherence time $t_{\rm D}$ for the model
            by considering
            the approximate solution to the master equation (\ref{master}),
            \[
             \rho_{\rm r}[\phi^+_<, \phi^-_<; t] \lesssim
            \rho^{\rm u}_{\rm r}[\phi^+_<, \phi^-_<; t] ~ \exp
            \bigg[-V\Gamma\int_0^t ds ~D_{\chi}( s) \bigg],
            \]
            where $\rho^{\rm u}_{\rm r}$ is the solution of the unitary part
            of the master equation (i.e. without environment). It is obvious
            from this (and also from (\ref{master})), that the diagonal
            density matrix just evolves like the unitary matrix (the
            environment has almost no effect on the diagonal part of
            $\rho_{\rm r}$). In terms of the dimensionless fields $\bar\phi =
            (\phi_<^+ + \phi_<^-)/2\mu,$ and $ \delta = (\phi_<^+ -
            \phi_<^-)/2\mu$, we have $\Gamma = (1/16)\bar g^2
            \mu^4\bar\phi^2\delta^2$.

            The system behaves classically when $\rho_{\rm r}$ is
            appropriately diagonal. We therefore look at the ratio
            \begin{equation}
            \left\vert \frac {\rho_{\rm r}[\bar\phi+\delta,\bar\phi-\delta;t]}
            {\rho_{\rm r}[\bar\phi,\bar\phi;t]}
            \right\vert \lesssim  \left\vert \frac {\rho_{\rm r}^{\rm u}
            [\bar\phi+\delta,
            \bar\phi-\delta;t]}
            {\rho_{\rm r}^{\rm u}[\bar\phi,\bar\phi;t]}
            \right\vert \exp
            \bigg[-V\Gamma\int_0^t ds ~D_{\chi}( s) \bigg ]\, .
            \label{ratio}\end{equation}

            It is not possible to obtain an analytic expression for the ratio
            of unitary density matrices that appears in Eq.(\ref{ratio}). The
            simplest approximation is to neglect the self-coupling of the
            system field \cite{guthpi}. In this case the unitary density
            matrix remains Gaussian at all times as

            \begin{equation}\left\vert \frac {\rho_{\rm r}^{\rm u}[\bar\phi+
            \delta,
            \bar\phi-\delta;t]}
            {\rho_{\rm r}^{\rm u}[\bar\phi,\bar\phi;t]}
            \right\vert = \exp [-{T_{\rm c}\over{\mu}}\delta^2 p^{-1}(t)]
            ,
            \label{uratio}
            \end{equation}
            where $p^{-1}(t)$, essentially $\mu^2
            \langle\phi^2\rangle_t^{-1}$, decreases exponentially with time to
            a value ${\cal O}(\lambda)$. This approximation can be improved by
            means of a Hartree-like approximation \cite{boya}. In this case
            the ratio is still given by Eq.(\ref{uratio}), but now $p^{-1}(t)$
            decreases more slowly as $t$ approaches $t_{\rm sp}$.  In any
            case, in the unitary part
            of the reduced density matrix the non-diagonal terms are not
            suppressed.  Therefore, in order to obtain classical behaviour, the
            relevant part of the reduced density matrix is the term
            proportional to the diffusion coefficient in Eq.(\ref{ratio}),
            since it is this that enforces its diagonalisation.

            The
            decoherence time  $t_{\rm D}$ sets the scale after which we have a
            classical system-field configuration.
            According to our previous discussion, it can be defined
            as the solution to
            \begin{equation}
            1\gtrsim V\Gamma \int_{0}^{t_{D}} ds ~D(
            s).
            \label{tchi}
            \end{equation}
            It corresponds to the time after which we are able to distinguish
            between two different field amplitudes (inside a given volume
            $V$).

            Suppose we reduce the couplings ${\bar g}$, $\lambda$ of the system
            $\phi$-field to its environment. Since, as a one-loop construct,
            $\Gamma\propto{\bar g}^2, \lambda^2$ our first guess would be
            that as ${\bar g}$, $\lambda$ decrease, then $t_D$ increases and
            the system takes longer to become classical. This is not really
            the case.
            The reason is twofold.
            Firstly, there is the effect that $\Gamma\propto T_0^2$, and
            $T_0^2 \propto\lambda^{-1}$ is non-perturbatively large for a
            phase transition.  Secondly, because of the non-linear coupling to
            the environment, obligatory for quantum field theory,
            $\Gamma\propto{\bar\phi}^2$. The completion of the transition
            finds ${\bar\phi}^2\simeq\eta^2\propto\lambda^{-1}$ also
            non-perturbatively large. This suggests that $\Gamma$, and hence
            $t_D$, can be independent of $\lambda$.  The situation would be
            different
            for a linear coupling to the environment,
            for which ${\bar\phi^2}$ would not be present, or a cold initial
            state in which $\phi$ is peaked about $\phi = 0$. In fact, the
            situation is a little more complicated, but the corollary that
            $t_D$ does not exceed $t_{\rm sp}$ as the couplings become weaker
            remains true.

            In order to quantify the decoherence time we have to fix
            the values of $V$, $\delta$, and $\bar\phi$. $V$ is understood as
            the minimal volume inside which we do not accept coherent
            superpositions of macroscopically distinguishable states for the
            field. Thus, our choice is that this volume factor is ${\cal
            O}(\mu^{-3})$  since $\mu^{-1}$ (the Compton wavelength) is the
            smallest scale at which we need to look. Inside this volume, we do
            not discriminate between field amplitudes which differ by $ {\cal
            O}(\mu) $, and therefore we take $\delta \sim {\cal O}(1)$.  For
            $\bar\phi$ we set $\bar\phi^2\sim {\cal O}(\alpha /\lambda)$,
             where $\lambda\leq\alpha\leq 1$ is to be determined
            self-consistently.

            Note that the diagonalisation of $\rho_t$ occurs in time as an
            {\it exponential} of an {\it exponential}. As a result,
            decoherence occurs extremely quickly, but not so quickly that $\mu
            t \ll 1$.  Consequently, in order to evaluate the decoherence time
            in our model, we have to use Eq.(\ref{Dt}). On taking the
            equality in
            (\ref{tchi}) we find
            \begin{equation}
            \exp [2\mu t_D] \approx {\lambda \sqrt{N}\bar g\over{\bar g^2
            \alpha}} = {\cal O}( \frac{\sqrt{N}}{\alpha}),
            \end{equation}
            whereby
            \[
            \mu t_D\simeq {1\over{4}}\ln N -{1\over{2}}\ln\alpha\simeq \ln
            (\frac{\eta}{T_{\rm c}\sqrt{\alpha}}).
            \]
            This is an {\it upper} bound on $t_D$, but probably
            still qualitatively correct.
            The value of $\alpha$ is determined as $\alpha \simeq
            \sqrt{\mu/T_{\rm c}}$ from the condition that, at time $t_D$,
            $\langle
            |\phi|^2\rangle_t\sim\alpha\eta^2$. Since $\alpha\ll 1$,
            in principle, the field has not diffused far from the
            top of the hill before it is behaving classically.

            For comparison, we find $t_{\rm sp}$, for which $\langle \phi^2
            \rangle_t \sim
            \eta^2$, given by
            \begin{equation}
            \exp [2\mu t_{\rm sp}] \approx  {\cal O}( \frac{\eta^2}{\mu
            T_{\rm c}}).
            \end{equation}
            The exponential factor, as always, arises from the growth of the
            unstable long-wavelength modes. The factor $T_{\rm c}^{-1}$ comes from
            the $\coth(\beta\omega /2)$ factor that encodes the initial
            Boltzmann distribution at temperature $T\gtrsim T_{\rm c}$. As a result,
            \begin{equation}
            \mu t_{\rm sp} \sim
            \ln (\frac{\eta}{\sqrt{\mu T_{\rm c}}}),
             \label{tsp}
            \end{equation}
            whereby $1 < \mu t_D \leq \mu t_{\rm sp}$, with
            \begin{equation}
            \mu t_{\rm sp} -\mu t_D\simeq \frac{1}{4}\ln (\frac{T_{\rm c}}
            {\mu})>1,
            \label{dt}
            \end{equation}
            for weak enough coupling, or high enough initial temperatures (we
            have taken $T_0 \sim T_{\rm c}$ throughout).
            This is our main result, that for the physically relevant modes
            (with small $k_0$) classical behaviour has been established before
            the spinodal time, when the ground states have became populated.

            \subsection{Back-reaction}

            We can now justify our earlier assumption that, for an
            instantaneous quench, nonlinear
            behaviour only becomes important just before the spinodal
            time\cite{Karra}. To see this, we adopt the
            Hartree approximation, in which the equations of motion are
            linearised so that the field modes $f^{\pm}_{k}$ now satisfy the
            equation
            \[
            \Biggl [ \frac{d^2}{dt^2} + {\bf k}^2 -\mu^2(t) \Biggr
            ]f^{\pm}_{k}(t) =0,
            \]
            where\cite{boya}
            \begin{equation}
            -\mu^2(t) = m^2(t) + 4\lambda\int {d^3p\over{(2\pi)^3}} \, C(p)
            [f^{+}_{p}(t)f^{-}_{p}(t)-1] \label{modeh2}
            \end{equation}
            and $C(p) = (\coth\beta\omega/2)/2\omega$, $\omega^2 = p^2 +m^2$.

            A similar result would be obtained by extending our initial $O(2)$
            theory to an $O(M)$ theory in the large-M limit before taking
            $M=2$.

            Our coarse-graining retains only the unstable modes in the
            integral, which suggests\cite{Karra} the approximate hybrid
            self-consistent equation for $\mu^{2}(t)$,
            \begin{equation}
            \mu^{2}(t)\simeq \mu^{2} - C\lambda \frac{T\mu}{(\mu
            t_{\rm sp})^{3/2}}\exp \bigg(2\int_{0}^{t}dt'\,\mu (t')\bigg),
            \label{muh2}
            \end{equation}
            ($C = O(1)$), which has the exponential growth of the WKB
            solution, but non-singular behaviour when $\mu (t)\approx 0$. The
            exact  solution to Eq.(\ref{muh2}) for $t<t_{\rm sp}$ is $\mu (t) =
            \mu\tanh \mu (t_{\rm sp}-t)$, irrespective of the values of the
            temperature $T$ and the coupling strength.
            In fact, we anticipated this in (\ref{tsp}), when we estimated
            $t_{\rm sp}$ on the assumption that the backreaction would only take
            effect in the final moments.

            That is, the theory only ceases to behave like a free Gaussian
            theory with upside-down potential at a time $t_{\rm B}$,
            \begin{equation}
            t_{\rm sp} - t_{\rm B} = O(\mu^{-1}). \label{tB}
            \end{equation}
            It follows that $t_{\rm B}\geq t_D$ in our ordering of scales
            $T_{\rm c}\gg\mu$ but, in practice $T_{\rm c}$ needs to be at least an order
            of magnitude larger than $\mu$ for this to be the case.

            When
            (\ref{dt}) is valid, we see that $\rho_{\rm r}$ becomes diagonal
            before non-linear terms could be relevant. In this sense,
            classical behaviour has been achieved before quantum effects could
            destroy the positivity of the Wigner function $W_{\rm r}$.

            \section{Final remarks}

            We have shown that, in our model, $\rho_{\rm r}$
            becomes diagonal before the spinodal time at which the order
            parameter field has first
            populated the ground-state values of the theory. Further, it can
            also be diagonal before
            non-linear terms are relevant. That is, decoherence can be
            achieved before quantum effects  destroy the positivity of the
            Wigner
            function $W_{\rm r}$. Really, our $t_{\rm D}$
            sets the time after which we have a classical probability
            distribution
            (positive definite) even for times $t > t_{\rm sp}$. The
            existence of the
            environment is crucial in doing this. Of course, for non-Gaussian
            or
            delocalised (in the field space) initial states, it is clear that
            $W_{\rm r}$
            will be non-positive definite even in the linear regime,
            and therefore
            $t_{\rm D}$ should be smaller than the one we evaluated here. In
            the present
            work, $t_{\rm D}$ is the {\it classicalisation time}-bound.

            This result goes in the direction of justifying the use
            of classical numerical simulations for the analysis of the
            dynamics of the long-wavelengths modes after the quench.

            \acknowledgments F.C.L. and F.D.M. were supported by Universidad
            de Buenos Aires, CONICET (Argentina),
            Fundaci\'on Antorchas, and ANPCyT. R.J.R. was supported in part
            by the COSLAB programme of the European Science
            Foundation. We also thank the organisers of the Peyresq VI meeting.

\end{document}